\documentclass[ALICE,manyauthors]{cernphprep}
\usepackage[comma,square,numbers,sort&compress]{natbib}
\usepackage{hyperref}
\usepackage{lineno}
\usepackage{xspace}
\usepackage{amsmath}
\usepackage{comment}
\usepackage{multirow}
\usepackage[T1]{fontenc}
\usepackage{orcidlink}
\usepackage{textcomp}
\usepackage[dvipsnames]{xcolor}
\usepackage{rotating}

\begin{document}
%

\newcommand{\pp}           {pp\xspace}
\newcommand{\ppbar}        {\mbox{$\mathrm {p\overline{p}}$}\xspace}
\newcommand{\XeXe}         {\mbox{Xe--Xe}\xspace}
\newcommand{\PbPb}         {\mbox{Pb--Pb}\xspace}
\newcommand{\pA}           {\mbox{pA}\xspace}
\newcommand{\pPb}          {\mbox{p--Pb}\xspace}
\newcommand{\AuAu}         {\mbox{Au--Au}\xspace}
\newcommand{\dAu}          {\mbox{d--Au}\xspace}

\newcommand{\s}            {\ensuremath{\sqrt{s}}\xspace}
\newcommand{\snn}          {\ensuremath{\sqrt{s_{\mathrm{NN}}}}\xspace}
\newcommand{\pt}           {\ensuremath{p_{\rm T}}\xspace}
\newcommand{\meanpt}       {$\langle p_{\mathrm{T}}\rangle$\xspace}
\newcommand{\ycms}         {\ensuremath{y_{\rm CMS}}\xspace}
\newcommand{\ylab}         {\ensuremath{y_{\rm lab}}\xspace}
\newcommand{\etarange}[1]  {\mbox{$\left | \eta \right |<#1$}}
\newcommand{\yrange}[1]    {\mbox{$\left | y \right |~<~#1$}}
\newcommand{\dndy}         {\ensuremath{\mathrm{d}N/\mathrm{d}y}\xspace}
\newcommand{\dndeta}       {\ensuremath{\mathrm{d}N_\mathrm{ch}/\mathrm{d}\eta}\xspace}
\newcommand{\avdndeta}     {\ensuremath{\langle\dndeta\rangle}\xspace}
\newcommand{\dNdy}         {\ensuremath{\mathrm{d}N_\mathrm{ch}/\mathrm{d}y}\xspace}
\newcommand{\Npart}        {\ensuremath{N_\mathrm{part}}\xspace}
\newcommand{\Ncoll}        {\ensuremath{N_\mathrm{coll}}\xspace}
\newcommand{\dEdx}         {\ensuremath{\textrm{d}E/\textrm{d}x}\xspace}
\newcommand{\RpPb}         {\ensuremath{R_{\rm pPb}}\xspace}
\newcommand{\etaranlr}[2] {\mbox{$#1<\eta<#2$}}
\newcommand{\acceff}       {\ensuremath{\mathrm{Acc \times Eff}\xspace}}
\newcommand{\multdensmid}  {\ensuremath{\langle\dndeta\rangle_{|\eta|~<~0.5}}\xspace}

\newcommand{\nineH}        {$\sqrt{s}~=~0.9$~Te\kern-.1emV\xspace}
\newcommand{\seven}        {$\sqrt{s}~=~7$~Te\kern-.1emV\xspace}
\newcommand{\thirteen}     {$\sqrt{s}~=~13$~Te\kern-.1emV\xspace}
\newcommand{\twoH}         {$\sqrt{s}~=~0.2$~Te\kern-.1emV\xspace}
\newcommand{\twosevensix}  {$\sqrt{s}~=~2.76$~Te\kern-.1emV\xspace}
\newcommand{\five}         {$\sqrt{s}~=~5.02$~Te\kern-.1emV\xspace}
\newcommand{\twosevensixnn}{$\sqrt{s_{\mathrm{NN}}}~=~2.76$~Te\kern-.1emV\xspace}
\newcommand{\fivenn}       {$\sqrt{s_{\mathrm{NN}}}~=~5.02$~Te\kern-.1emV\xspace}
\newcommand{\LT}           {L{\'e}vy-Tsallis\xspace}
\newcommand{\GeVc}         {Ge\kern-.1emV/$c$\xspace}
\newcommand{\MeVc}         {Me\kern-.1emV/$c$\xspace}
\newcommand{\TeV}          {Te\kern-.1emV\xspace}
\newcommand{\GeV}          {Ge\kern-.1emV\xspace}
\newcommand{\MeV}          {Me\kern-.1emV\xspace}
\newcommand{\GeVmass}      {Ge\kern-.2emV/$c^2$\xspace}
\newcommand{\MeVmass}      {Me\kern-.2emV/$c^2$\xspace}
\newcommand{\lumi}         {\ensuremath{\mathcal{L}}\xspace}
\newcommand{\degree}       {\ensuremath{^{\rm o}}\xspace}

\newcommand{\ITS}          {\rm{ITS}\xspace}
\newcommand{\TOF}          {\rm{TOF}\xspace}
\newcommand{\ZDC}          {\rm{ZDC}\xspace}
\newcommand{\ZDCs}         {\rm{ZDCs}\xspace}
\newcommand{\ZNA}          {\rm{ZNA}\xspace}
\newcommand{\ZNC}          {\rm{ZNC}\xspace}
\newcommand{\SPD}          {\rm{SPD}\xspace}
\newcommand{\SDD}          {\rm{SDD}\xspace}
\newcommand{\SSD}          {\rm{SSD}\xspace}
\newcommand{\TPC}          {\rm{TPC}\xspace}
\newcommand{\TRD}          {\rm{TRD}\xspace}
\newcommand{\VZERO}        {\rm{V0}\xspace}
\newcommand{\VZEROA}       {\rm{V0A}\xspace}
\newcommand{\VZEROC}       {\rm{V0C}\xspace}
\newcommand{\Vdecay} 	   {\ensuremath{\rm{V^0}}\xspace}

\newcommand{\ee}           {\ensuremath{e^{+}e^{-}}} 
\newcommand{\pip}          {\ensuremath{\pi^{+}}\xspace}
\newcommand{\pim}          {\ensuremath{\pi^{-}}\xspace}
\newcommand{\kap}          {\ensuremath{\rm{K}^{+}}\xspace}
\newcommand{\kam}          {\ensuremath{\rm{K}^{-}}\xspace}
\newcommand{\pbar}         {\ensuremath{\rm\overline{p}}\xspace}
\newcommand{\kzero}        {\ensuremath{{\rm K}^{0}_{\rm{S}}}\xspace}
\newcommand{\lmb}          {\ensuremath{\Lambda}\xspace}
\newcommand{\almb}         {\ensuremath{\overline{\Lambda}}\xspace}
\newcommand{\Om}           {\ensuremath{\Omega^-}\xspace}
\newcommand{\Mo}           {\ensuremath{\overline{\Omega}^+}\xspace}
\newcommand{\X}            {\ensuremath{\Xi^-}\xspace}
\newcommand{\Ix}           {\ensuremath{\overline{\Xi}^+}\xspace}
\newcommand{\Xis}          {\ensuremath{\Xi^{\pm}}\xspace}
\newcommand{\Oms}          {\ensuremath{\Omega^{\pm}}\xspace}
\newcommand{\Xit}          {\ensuremath{\Xi}\xspace}
\newcommand{\Omt}          {\ensuremath{\Omega}\xspace}
\newcommand{\Lamboth}      {\ensuremath{\Lambda\kern-.1em+\overline{\Lambda}}\xspace}
\newcommand{\Xiboth}       {\ensuremath{\Xi^-\kern-.4em+\overline{\Xi}^+}\xspace}
\newcommand{\Omboth}       {\ensuremath{\Omega^-\kern-.4em+\overline{\Omega}^+}\xspace}

\newcommand \abs[1]{\left|#1\right|}

\newcommand{\pem}       {Pythia 8 Monash\xspace}
\newcommand{\pecrr}     {Pythia 8 QCD-CR Ropes\xspace}
\newcommand{\eplhc}     {EPOS LHC\xspace}
\newcommand{\epf}       {EPOS 4\xspace}

\begin{titlepage}

\PHyear{2025}       
\PHnumber{254}      
\PHdate{04 November}  

\title{Centrality dependence of strange particle production in Pb--Pb collisions at $\mathbf{{\sqrt{s_{\mathrm{NN}}}~=~5.02~Te\kern-.1emV}}$}
\ShortTitle{Strangeness production in Pb--Pb collisions at \fivenn}

\Collaboration{ALICE Collaboration\thanks{See Appendix~\ref{app:collab} for the list of collaboration members}}
\ShortAuthor{ALICE Collaboration} 

\begin{abstract}
The centrality dependence of strange (\kzero, \Lamboth) and multi-strange (\Xiboth, \Omboth) hadron production is measured by ALICE in the LHC lead--lead (Pb--Pb) collisions at a center-of-mass energy per nucleon pair \fivenn, using the full data set collected during the LHC Run 2 campaign in the years 2015 and 2018. This is the largest heavy--ion data set analyzed to date at the LHC, and it allows for the extraction of transverse momentum (\pt) spectra and \pt-integrated yields with unprecedented precision, over a broad range of charged particle multiplicity densities (\multdensmid), probing regions where smaller collision system (pp and p--Pb) results are also available. 
The \pt spectra evolve with centrality, featuring higher \meanpt in central events for all particles. The \lmb/\kzero ratio exhibits the distinctive baryon-to-meson enhancement in the intermediate \pt region, with a maximum which is shifted to larger \pt for more central collisions. The hadron-to-pion yield ratios are presented as a function of \multdensmid and compared to results from different collision systems and energies. A smooth connection from \pp to \PbPb is observed, thus demonstrating that collision system or energy do not play a role in the multiplicity evolution of this observable. The previously reported enhancement of strangeness production in the multiplicity range probed in \pp and \pPb collisions saturates in the multiplicity range of \PbPb data.
These results constitute a key test bench for theoretical models and a first comparison to the EPOS~4 generator is presented.

\end{abstract}
\end{titlepage}

\setcounter{page}{2}

\section{Introduction}
Strange hadron production studies in ultra-relativistic heavy-ion collisions are an important tool to characterize the strongly interacting matter known as quark--gluon plasma (QGP). As predicted by quantum chromodynamics (QCD), this state is created when nuclear matter, under extreme energy density conditions ($\epsilon \gtrsim \mathrm{0.5~GeV/fm^3}$~\cite{lqcdepsilon}), undergoes a transition into a system whose relevant degrees of freedom are partonic. In this respect, a modified (multi-)strange hadron production probability was originally proposed as a signature for QGP formation~\cite{Rafelski:1980, Rafelski:1986, KOCH1986167}.
The argument was that in the high-temperature deconfined state of strongly-interacting matter, the abundant presence of gluonic excitations would cause a higher rate of strange quark--antiquark production, leading to a fast equilibration of strangeness. In contrast to a purely hadronic gas scenario, where reaction rates are too low to allow for equilibration of strange particle densities, QGP formation would lead to an enhancement of strange hadrons and, in particular, of multi-strange (anti)hyperons. \\
However, it has also been argued that strangeness enhancement strongly depends on the thermodynamical circumstances considered. Therefore, the predicted enhancement due to the plasma might not be as large as previously expected~\cite{Redlich:1985,Cleymans:1991}. The comparison of a plasma with a hadronic system in equilibrium with the same amount of entropy and baryon number showed that strangeness could be equally abundant in both cases~\cite{PhysRevC.37.1452}. In fact, Statistical Hadronization Models (SHM) with their Grand Canonical formalism have been successful in fitting the abundance of light hadrons over a wide range of masses, from pions to light nuclei, and across various energies~\cite{Becattini:therm97,Anton:2006,Anton:2018,Flor:2020fdw,Fernando:22}. In this approach, the baseline for strange hadron production is the measurement in heavy-ion collisions. However, the smaller abundances observed in hadronic interactions could be regarded as coming from the limited phase space available due to local strange quantum number conservation (the so-called ``canonical suppression''~\cite{Tounsi:2001ck}). Moreover, in small-volume systems, the $\gamma_{\mathrm{S}}$ approach~\cite{Becattini:06,Volodya:19} accounts for the fact that strangeness may not reach full chemical equilibrium, meaning that its abundance would not match the one expected in a fully thermalized scenario.
In a scenario without QGP formation, but considering alternate processes like hadronization through string formation in multiple collisions, strangeness production could be significantly enhanced as well~\cite{Sorge:1992,Sorge:1995,Bierlich:2015}.

The comparison between strange hadron production in heavy-ion and smaller collision systems led to the observation of the strangeness enhancement phenomenon, from the energy ranges of the CERN Super Proton Synchrotron (SPS) with $\sqrt{s_{\mathrm{NN}}}~=~17.3$ \GeV~\cite{NA35:se,WA97:se99,WA85:se99,Antinori:2002qr,NA57:se06}, through the Relativistic Heavy Ion Collider (RHIC) with $\sqrt{s_{\mathrm{NN}}}~=~200$ \GeV~\cite{STAR:se08,STAR:se12} up to the LHC at \twosevensixnn~\cite{ALICE_cascPbPb276}. Furthermore, in the energy range from RHIC to LHC, the hyperon-to-pion ratio as a function of collision centrality exhibits a universal behavior across different collision energies, showing a feeble increase of the ratio from peripheral to central collisions and reaching a saturation value from mid-to-central collisions.
In recent years, the extension of these studies to smaller collision systems such as \pp and \pPb triggered large interest in both particle and high energy nuclear physics theory communities. The hyperon-to-pion ratio was studied as a function of event multiplicity instead of using geometrical quantities, allowing for a direct comparison of results obtained in different collision systems. Surprisingly, strangeness enhancement was observed in high-multiplicity \pp~\cite{Alice_NPEnhProd,ALICEstrPP7,ALICEstrPP13,silviapp5_TBU} and \pPb~\cite{Alice_strpPb5,ALICE_cascpPb5} interactions, with a pattern that is insensitive to the collision center-of-mass energy and system. To solidify this observation, significant emphasis has been given to extending the study to the highest accessible multiplicity in small systems. However, the extension to the most peripheral collisions in \PbPb is equally important in order to cover a multiplicity region common to all the three systems. The latter is the focus of the work reported here.

In this paper, results on strange (\kzero, \lmb and \almb) and multi-strange (\X, \Ix, \Om and \Mo) hadron production in \PbPb collisions at \fivenn are presented.
These results constitute the most precise determination of strange particle transverse momentum (\pt) distributions and yields at the highest center-of-mass collision energy provided by LHC during its Run 2 (2015--2018) \PbPb data taking period. As such, they will serve as a key test bench for a precise determination of the thermodynamic parameters of the created system and for the tuning of Monte Carlo (MC) event generators.
A comprehensive model comparison is out of the scope of this paper, but a first comparison of the results to the \epf MC event generator~\cite{EPOS4:factosatu, EPOS4:pqcd} is carried out. EPOS is based on the Gribov--Regge theory at the parton level~\cite{EPOS:Gribov}, with an hydrodynamic evolution and statistical hadronization of the hot regions (core) and in-vacuum hadronization of the colder regions (corona). In version 4 of the EPOS model, considerable focus has been placed on tuning the dynamics associated with strange particle formation ~\cite{EPOS4:strangeness}, making it a suitable tool to highlight all the general features connected to the hadronization of a hot extended rapidly expanding system, from hydrodynamic parton flow, to quark coalescence and gluon fragmentation dynamics. At the same time, it enables a comparison of the expected yields from its core–corona approach with the real data.

The paper is organized as follows: Section 2 summarizes the ALICE setup used during the LHC Run 2 campaign, while Section 3 provides details on the data analysis, including sample definition and all steps from reconstruction to signal extraction and corrections. In the same section, a description of all sources of systematic uncertainty is given. Finally, Section~4 presents the results and Section~5 draws the conclusions.

\section{Experimental setup}
A detailed description of the ALICE detector and its performance can be found in Ref.~\cite{Aamodt:2008, ALICE:2014sbx}. 
Strange hadrons are measured in ALICE through the reconstruction of their decay daughter particles, arranged in a distinctive weak-decay topology. 
The relevant detectors for this analysis can be divided into two sub-groups: the midrapidity tracking and particle identification (PID) systems and the forward rapidity detectors used to trigger and select events. 
The ALICE central barrel tracking system consists of the ITS (Inner Tracking System)~\cite{Aamodt:2010} and the TPC (Time Projection Chamber)~\cite{Alme:2010}. 
The ITS detector covers the pseudorapidity range $|\eta| < 0.9$ and it is further divided into three different subdetectors, each consisting of two cylindrical layers surrounding the beam vacuum tube. The four outer layers are based on silicon drift and double-sided silicon strip technologies, used for tracking and particle identification. The  innermost two layers, based on hybrid-pixel sensors, are used for tracking and for primary vertex determination.
The main tracking detector is the TPC, which provides full azimuthal and a longitudinal coverage of \etarange{0.9}, allowing up to 159 reconstructed clusters per track. The TPC also enables precise particle identification through specific ionization energy loss in the detector gas (\dEdx). 
The outermost detector used in this work is the Time Of Flight (TOF)~\cite{TOFdet}, positioned at a radial distance of 370~cm from the beam axis. It has a pseudorapidity coverage of $|\eta| < 0.9$, full azimuthal acceptance, and it is used to measure the time-of-flight of particles with an intrinsic time resolution of < 60 ps.
 A solenoidal magnet generates a uniform 0.5 T field along the longitudinal axis, bending charged particle tracks in the transverse plane.
The main detector used for triggering is the V0 system~\cite{Abbas:2013}, an array of scintillator detectors comprising two parts placed in the forward (\etaranlr{2.8}{5.1}) and backward (\etaranlr{-3.7}{-1.7}) regions outside the TPC coverage. 
Event selection also uses data from the Zero Degree Calorimeter (ZDC)~\cite{CERN-LHCC-99-005}, located at $\rm{\pm112.5~m}$ from the interaction point (IP), to reject electromagnetic interactions and to remove background from collisions involving charge circulating outside the nominal LHC bunch position.
Additionally, the signal amplitude in the V0 detector is employed to determine collision centrality~\cite{ALICEcentrality}, expressed in terms of the percentile of the hadronic cross section, with 0\% identifying the most central collisions.
\section{Data analysis}
\subsection{Data sample and event selection}
\label{sec:data}
This analysis is based on 280 million \PbPb collision events at \fivenn, comprising the entire \PbPb data sample collected in 2015 and 2018 by ALICE during the LHC Run 2 campaign. On top of a minimum-bias trigger, requiring at least one hit in both the V0 detectors, two specific triggers were implemented, called central and semi-central, where the thresholds of the V0 detector signal were tuned in order to collect additional events in the 0--10\% and 30--50\% centrality intervals.
Only collisions occurring within 10~cm from the nominal IP position in the longitudinal direction are accepted to ensure uniform acceptance within the measured $\eta$ range and to avoid border effects. Beam-induced background events are removed offline exploiting the V0 and ZDC timing information~\cite{Adam:2015}. Pileup of collisions from different bunch crossings in the TPC is rejected by comparing multiplicity estimates from the V0 detector to those of tracking detectors at midrapidity, exploiting the difference in readout times between the systems.
The data sample is divided into nine equally-wide centrality intervals, ranging from 0$\%$ to 90$\%$. The 90-100$\%$ centrality class is not considered for the analysis because it is significantly contaminated by electromagnetic processes.

\subsection{Strange particle reconstruction and corrections}
Strange and multi-strange hadrons are detected by reconstructing their weak decay topologies~\cite{pdg}:
\begin{align*}
    &\kzero  \rightarrow \pip + \pim                   & &\mbox{B.R.} = (69.20 \pm 0.05) \% \\ 
    &\lmb    \rightarrow \mathrm{p} + \pim     \mbox{(+ c.c.)}  & &\mbox{B.R.} = (64.1 \pm 0.5) \% \\
    &\X      \rightarrow \lmb + \pim  \mbox{(+ c.c.)}  & & \mbox{B.R.} = (99.887 \pm 0.035) \% \\ 
    &\Om     \rightarrow \lmb + \kam  \mbox{(+ c.c.)}  & & \mbox{B.R.} = (67.8 \pm 0.7) \%
\end{align*}
In the rest of the paper, when similar considerations apply to particles and antiparticles, the generic symbols \lmb, \Xit and \Omt will be used.
Given that \kzero and \lmb are charge neutral, the decay is characterized by a V-shape topology (hence the name of \Vdecay particles) and it is built by pairing opposite-charge tracks reconstructed in ITS and TPC.
Multi-strange baryons decay into a pair formed by a \Vdecay and a charged particle, also known as bachelor, and it is followed by the weak decay of the \Vdecay as described before (hence the name of cascade particles). The reconstruction process relies on the standard ALICE weak-decay finder~\cite{ALICE:strpp900}, which implements very loose track quality, kinematical, and geometrical selections. These criteria are then tightened at the analysis level to reduce the large combinatorial background. The list of final selections used in this analysis is reported in Table~\ref{tab:selections}. 

\begin{table}[h]
  \caption{Selection criteria applied to the (multi-)strange candidates and their decay products. When the selection criterion value changes between \kzero and \lmb or between \Xit and \Omt, the values for \lmb and \Omt are given in brackets. The topological selection marked with $^{(*)}$ is applied only in the 0--40\% centrality range, and it is parameterized across \pt, ranging between the low- and high-\pt values separated by $\rightsquigarrow$ (more information in the text).}
  \centering
  \begin{tabular}{cc}
    \hline
    daughter-track selection criterion     &  value        \\
    \hline
    $|\eta|$                                & $<0.8$        \\
    TPC crossed rows                            & $\ge80$       \\
    TPC crossed rows over findable              & $\ge0.8$      \\
    $\chi^2$ per TPC cluster                & $\le2.5$      \\
    TPC n$_\sigma$                          & $\le5$        \\
    \hline
    \Vdecay candidate selection criterion      & value             \\
    \hline
    $\cos\theta_\mathrm{pa}$                & $>0.97$($>0.99$)  \\ 
    $R_{2D}$~(cm)                             & $1-125$        \\
    DCA \Vdecay daughters--PV~(cm)            & $>0.11$         \\
    DCA \Vdecay daughters~($\sigma$)          & $<1$          \\  
    $\pt^\mathrm{Arm}$ (GeV/$c$)                     & $>0.13$ (not applied)         \\
    Proper lifetime~($n_{\rm c\tau}$)               & $<7.5$~($<3.8$)            \\   
    $|y|$                               & $<0.5$            \\
    $n_{\rm daughters}$ with ITS refit or hit in TOF & $\geq1$               \\ 
    \hline
    Cascade candidate selection criterion    & value     \\
    \hline
    $\cos\theta_\mathrm{pa}$                & $>0.99$        \\
    \Vdecay $\cos\theta_\mathrm{pa}$          & $>0.99$       \\ 
    $^{(*)}$ bachelor--baryon $\theta_\mathrm{pa}$ & $>2.56^{\circ} \rightsquigarrow 0.08^{\circ}$ \\ 
    cascade $R_{2D}$~(cm)                        &  $>1$       \\
    \Vdecay $R_{2D}$~(cm)                          & $3-85$      \\
    DCA \Vdecay--PV~(cm)                      & $>0.1$           \\
    DCA bachelor--PV~(cm)                   & $>0.1$           \\
    DCA \Vdecay daughters--PV~(cm)            & $>0.2$            \\
    DCA bachelor--\Vdecay~(cm)                & $<1$            \\
    DCA \Vdecay daughters~($\sigma$)          & $<1$                \\  
    Proper lifetime~($n_{\rm c\tau}$)               & $<3.0$           \\   
    \Vdecay mass window~(MeV/$c^2$)           & $\pm 5$                 \\
    Competing mass rejection~(MeV/$c^2$)             & not applied ~($\pm 8$)               \\
    $|y|$                               & $<0.5$         \\
    \hline
  \end{tabular}
  \label{tab:selections}
\end{table}

The most effective geometrical selection on the displaced decay-vertex topology is related to the pointing angle ($\theta_\mathrm{pa}$), the angle between the reconstructed mother particle momentum and the straight line connecting the primary vertex and decay vertices. A tight selection on the cosine of the pointing angle ($\cos\theta_\mathrm{pa}$) for all the weak decays involved in the analysis guarantees the decay products pointing towards their decay vertex. A pointing angle selection is also applied in the analysis of multi-strange particles to exclude candidates built-up from a true \lmb particle, whose meson daughter is retained as the cascade's bachelor and whose baryon daughter is wrongly associated to an unrelated pion. In this case, a diversified selection on the minimum bachelor-to-baryon pointing angle is developed to remove wrong candidates, whose presence appears as a structure in the cascade's invariant mass distribution in central events and at low transverse momentum, with selection values going from $2.56^{\circ}$ at $\pt\approx0.8$~GeV/$c$ to $0.08^{\circ}$ at $\pt\approx3.2$~GeV/$c$. Since moving from central to peripheral events the loss of candidates increases and the impact of the wrong topology association becomes less important, this selection is applied in the centrality range 0--40\% only. Additional selections on the decay radius ($R$), defined in the plane perpendicular to the beam direction, are set in order to reject decays not displaced enough or too far from the interaction point, to guarantee an accurate secondary vertex determination and good TPC tracking performance for the decay products. Reconstructed tracks are propagated towards the interaction point in order to compute the distance of closest approach (DCA) with respect to the primary vertex (PV) or between pairs of tracks. Candidates are rejected depending on DCA selections which limit the presence of decay topologies not compliant with displaced secondary vertices or with decay products not pointing to the same space origin. Proper lifetime is also considered as a criterion to exclude candidates which decay too far from the IP with respect to the expected $c\tau$ for the particles under study. To suppress the contamination of $\lmb$ baryons reconstructed as $\kzero$, a selection on the maximum allowed Armenteros--Podolanski transverse momentum ($\pt^\mathrm{Arm}$), i.e. the transverse momentum of the positive \Vdecay daughter with respect to the \Vdecay momentum vector, is applied. In the cascade analysis, a selection on the estimated invariant mass of the \Vdecay decay candidate is performed, requiring that it does not fall too far from the expected nominal \lmb mass~\cite{pdg}. $\Omt$ candidates whose invariant mass computed under the competing cascade hypothesis falls in a close window around the nominal \Xit mass~\cite{pdg} are discarded to avoid large contamination from wrongly identified decay daughters. In addition to the selections related to the decay topology, a set of kinematic and track quality selections is also implemented. A first requirement on the particle rapidity and on its daughters' pseudorapidity is introduced to consider candidates well inside the central barrel region, in order to ensure a uniform tracking performance. Good quality TPC tracks are also required in terms of minimum number of TPC crossed raws and good TPC track fit quality. In addition, a successful refit of TPC tracks at the end of the reconstruction process is also required. Particle identification selections are performed on all decay products with the TPC detector, retaining daughter tracks whose energy loss in the detector gas is within a given range of $n_\sigma$ from the expected \dEdx, where $\sigma$ is the \dEdx resolution. In addition, a dedicated selection is applied to remove those tracks falling for a large fraction of their path in the dead areas between TPC azimuthal sectors or in pads close to the sector boundaries of the TPC detector. These tracks are not correctly modeled in the simulations, so that their removal allows avoiding biases in the analysis and to have better track quality. This selection is fully applied in the \Vdecay analysis, while it is used only up to a centrality of 40\% for multi-strange particles, since no effects on the corrected spectra for peripheral events are observed in this case. Finally, in order to reduce the effect of candidates coming from out-of-bunch pile-up events not rejected at the event selection step, for the \Vdecay analysis it is required that at least one of the daughter tracks is refitted within the ITS detector or is associated to a cluster in the TOF detector. This selection is not required for multi-strange particles, since, despite a severe reduction in statistics, full compatibility is observed in the final results with and without its application.
 
\begin{figure}[ht]
	\centering
    \includegraphics[width=\textwidth]{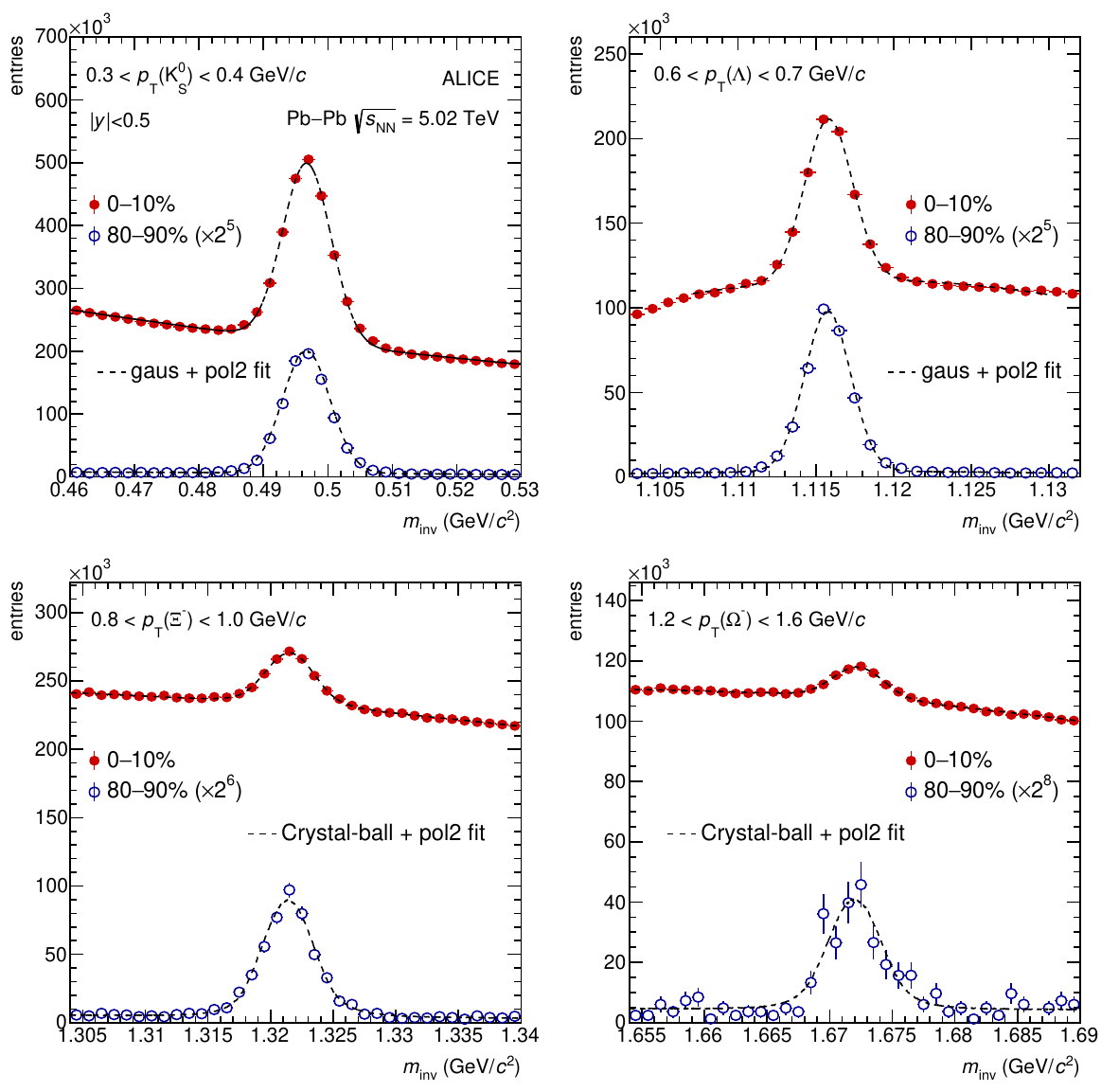 }
     \caption{Invariant mass distributions for (top left) $\kzero$ (top right) $\lmb$ (bottom left) $\X$ and (bottom right) $\Om$ for the lowest transverse momentum intervals accessed by the analysis. The distributions for the most central (red full markers) and most peripheral (blue open markers) class of events are reported. The dashed lines illustrate the fits used for signal extraction.}
     \label{fig:invMass}
\end{figure}
After all selections, candidates are divided in \pt and centrality intervals and the invariant mass is computed. Examples of the invariant mass distribution for strange and multi-strange particles are reported in the top and bottom panels of Fig.~\ref{fig:invMass}, for the lowest transverse momentum region in which a successful signal extraction is possible. In general, the signal-over-background ratio improves with increasing \pt, so the figure shows the situation where the signal extraction is the most challenging due to the large combinatorial background. The signal-over-background ratio increases from central to peripheral collisions. In order to extract the raw yields, a fit to the invariant mass distribution is performed. The fit function is modeled with the sum of two functions which describe the signal and the combinatorial background respectively. While the \Vdecay signal is described by a Gaussian shape, for multi-strange particles a Crystal-Ball function~\cite{Gaiser:1982yw} is used. In the latter case, the non-Gaussian tails are fixed by fitting the invariant mass distribution of pure signal candidates extracted from MC simulations in each \pt and centrality interval. The combinatorial background is modeled with a second (pol2) or first-degree (pol1) polynomial function, depending on the particle species, centrality, and transverse momentum. The raw yield is then estimated with the bin counting method, i.e. by subtracting the integral of the background function from the total number of candidates in the peak region. The peak region is defined as a $\pm n\sigma$ invariant mass window around the signal peak position. For \Vdecay, where the signal peak is narrow and non-Gaussian tails do not contribute significantly to the yield estimation, $n$ is fixed to 3, while for cascades $n=5$.

\begin{figure}[ht]
	\centering
    \includegraphics[width=\textwidth]{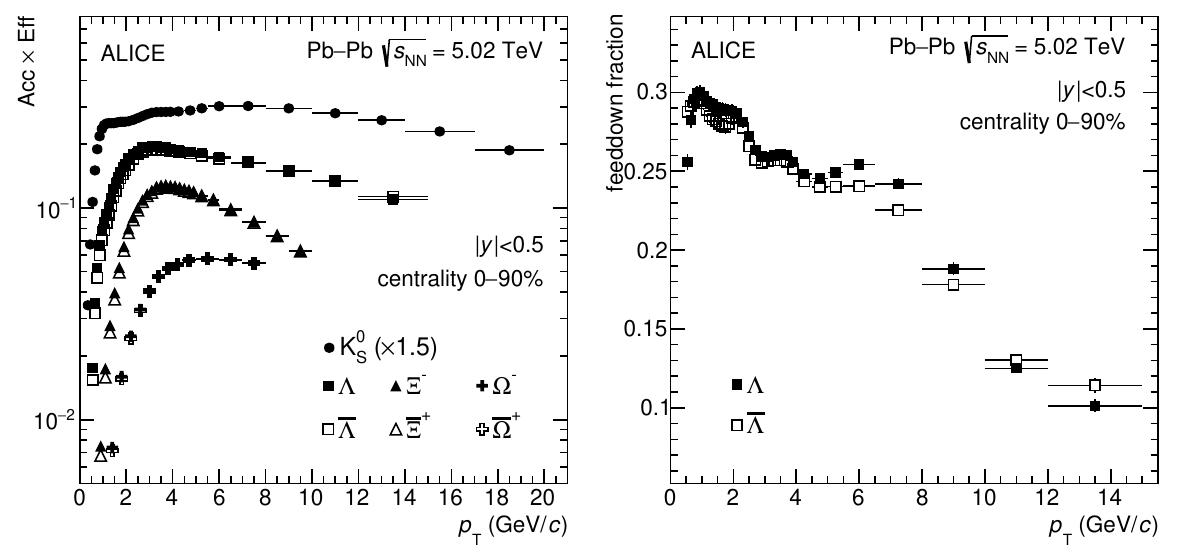}
     \caption{(left) Acceptance times efficiency (\acceff) as a function of \pt for the reconstruction of \kzero, \lmb, \X, \Om (full markers) and respective antiparticles (open markers). The 0--90\% centrality class is displayed. (right) Ratio of \lmb coming from charged and neutral \Xit baryon decays to the inclusive \lmb yield as a function of \pt. Full and hollow markers are for matter and antimatter, respectively. The 0--90\% centrality class is displayed.}
     \label{fig:accefffdw}
\end{figure}
A correction factor is considered to take into account the detector acceptance, the track reconstruction efficiency, and the efficiency of the applied candidate selection criteria (\acceff). This correction is evaluated from MC simulation of HIJING~\cite{Gyulassy:1994ew} events injected with multiple \Vdecay and cascade particles in each event to achieve negligible statistical uncertainty with respect to real data. The generated particles are fed to a transport code, which can be either GEANT3~\cite{Brun:1994aa} or GEANT4~\cite{GEANT4:2002zbu}. The comparison between two MC samples obtained with the different transport codes shows compatible results, so the two are merged. Figure~\ref{fig:accefffdw}~(left) shows the \acceff~correction factor integrated in the centrality interval 0--90\%, including the branching ratio factor, as a function of \pt for all particles under study. \acceff~corrections are estimated for each centrality class, as large multiplicities and tightened selections implemented in central collisions have a significant effect on \acceff.
For the $\lmb$ analysis, an additional correction to remove feed-down candidates from \Xit decays is applied to the raw spectra: a 2D matrix describing the decay kinematics and incorporating secondary \lmb reconstruction efficiencies is weighted for twice the measured \Xit spectrum (to account for the charged and neutral cascade decays). Figure~\ref{fig:accefffdw}~(right) reports the estimated fraction (in the 0--90\% centrality range) of uncorrected \lmb candidates which come from the decay of \Xit particles. The fraction is strongly \pt dependent and reaches a maximum of 30\% in the low-\pt region. The observed \pt modulation reflects the slightly different kinematics of primary and secondary \lmb and it becomes visible where the \pt bin width changes (e.g. at 2 and 4 \GeVc). As previously discussed for \acceff, also in this case, the fraction is centrality dependent, reaching its highest value at intermediate \pt in central collisions (35\%) and decreasing to 23\% in the 80--90\% centrality class. 

Integrated yields and \meanpt are obtained by extrapolating the \pt spectra in the low-\pt region by means of a fit with individual Blast-Wave functions~\cite{art:blastwave}. Other parametrizations have been exploited to assess the sensitivity to the choice of the functional form on the extrapolated fraction, as discussed in the next paragraph.

\subsection{Systematic uncertainties} 
The list of all sources of systematic uncertainty for the 0--90\% centrality class is summarized in Table~\ref{tab:syst} for two representative \pt intervals at low and high \pt. For all baryons, the values corresponding to the sum of particles and antiparticles are reported.
The major contribution to the total systematic uncertainty comes from the adopted track and candidate selection criteria. This uncertainty is evaluated by changing each selection with respect to the default value in a range leading to a $\pm 10\%$ variation in the raw yields and re-evaluating the corrected spectra. The maximum variation in the final results for each  selection variable is then quadratically added to all other contributions. The estimation of the systematic uncertainty related to the signal extraction is different among the considered species. For both strange and multi-strange particles, the interval in which the counting of candidates is performed is varied by $\pm 1\sigma$ with respect to the default case. Additionally, for $\kzero$ and $\lmb$ particles, the functions exploited in the signal extraction are varied: the raw signal is re-evaluated by using either a double Gaussian for the signal or a third-degree polynomial for the background. This approach is not suitable for multi-strange particles due to the more complicated shape of the invariant mass distributions. In this case, a modified background is considered by varying the interval in which the invariant mass distributions are fitted. Moreover, the integral of the signal function is used as raw signal evaluation instead of the default bin counting technique. 
Another source of systematic uncertainty is the implementation of the material budget of the detector in the MC simulations. This is evaluated by comparing different MC simulations in which the material budget of the ALICE detector was varied by $\pm$4.5\%, corresponding to the uncertainty in the determination of the material budget obtained by measuring photon conversions~\cite{ALICE:2014sbx}. The effect of the out-of-bunch pile-up rejection is also tested by changing the number of daughter tracks that are required to be ITS refitted or associated to a TOF hit. Another contribution is considered to account for the track quality for the selection on the fraction of track path in the insensitive regions between TPC sectors, whose parametrization is varied to study its effect on the final corrected yields. 
The effect of secondary (multi-)strange particles coming from interactions with the beam pipe or the detector material is also studied via the MC simulation, and it leads to negligible uncertainty. In the \lmb analysis, the feed-down estimation is varied by shifting the \Xit spectra by $\pm 1\sigma$ of their systematic uncertainty, thus accounting for the imprecision in the cascade spectrum determination. Finally, since the overall data sample used for the analysis consists of data collected in two different periods, a contribution taking into account the difference in corrected yields between the 2015 and 2018 data samples is considered. This amounts to an effect smaller than 3\%, thus showing the stability of the ALICE detector and running conditions during the entire LHC Run 2 campaign. 

\begin{table}[h]
\centering
\caption{Summary of systematic uncertainties (classified by source) in percent on the \kzero, \lmb, \Xit, and \Omt \pt spectra measurement for the 0--90\% centrality class. Values at low and high \pt are displayed.}
\label{tab:syst}
\renewcommand{\arraystretch}{1.2}
\begin{tabular}{lcccccccc}
\hline
                   & \multicolumn{2}{c}{$\kzero$} & \multicolumn{2}{c}{$\lmb$} & \multicolumn{2}{c}{$\Xit$} & \multicolumn{2}{c}{$\Omt$} \\
$p_\mathrm{T}$ (GeV/$c$)           &  0.35  &  18.5    &  0.65  &  13.5  &  0.9    &  9.5    &  1.4    &  7.5    \\ \hline
Track and candidate selection       &  3\%  &  4\%     &  13\%  &  12\%   &  6\%    &  15\%   &  11\%   &  17\%   \\
Signal extraction                  &  6\%  &   10\%   &  3\%  &  6\%   &  1\%    &  4\%    &  2\%    &  7\%    \\
Material budget                    &  1\%  &  <1\%    &  4\%  &  <1\%   &  4\%    &  1\%    &  3\%    &  1\%    \\
Pile-up rejection                  &  <1\%  &  2\%    &  <1\%  &  5\%   &  < 1\%  &  2\%    &  < 1\%  &  8\%    \\
Track-length in TPC inactive areas               &  <1\%  &  <1\%   &  3\%  &  7\%   &  4\%    &  6\%    &  5\%    &  12\%   \\
Secondary particles from material  &  <1\%  &  <1\%   &  1\%  &  <1\%   &  < 1\%  &  < 1\%  &  < 1\%  &  < 1\%  \\
Feeddown Correction                & \--    & \--     & 4\% & 2\% & \-- & \-- & \-- & \-- \\
Consistency between data samples              &  1\%  &  1\%     &  3\%  &  2\%   &  < 1\%  &  3\%    &  2\%    &  2\%    \\ 
 \hline
Total                              &  7\%  &  11\%    &  15\%  &  15\%   &  8\%    &  17\%   &  13\%   &  23\%   \\ \hline
\end{tabular}
\end{table}

The effect of different \pt parametrizations in the extraction of integrated yields and \meanpt is studied by exploring several functional forms in the low-\pt extrapolation procedure, such as L\'evy--Tsallis~\cite{Tsallis:1988}, Bylinkin~\cite{art:Bylinkin},  Hagedorn~\cite{art:Hagedorn}, Bose--Einstein, and Fermi--Dirac. This source of systematic uncertainty is centrality dependent, being the highest in peripheral collisions (6\%, 5\%, 4\%, and 6\% for \kzero, \lmb, \Xit, and \Omt, respectively, in the 80--90\% centrality class) and the lowest in central ones (2\%, 1\%, 1\%, and 3\%, respectively, in the 0--10\% centrality class). For the \meanpt measurement, the systematic uncertainty associated to the extrapolation function is in general lower, reaching the highest value of 5\% for peripheral collisions in the \Omt case.

The normalisation uncertainty due to the centrality interval definition was estimated by varying the limits of the centrality classes to account for the uncertainty on the fraction of the hadronic cross section used in the Glauber fit to the total amplitude from the V0 detector~\cite{ALICE:2015vxz}.
It reaches 10\% in the most peripheral event class, while it decreases down to <0.5\% in the 0--10\% class.
The systematic uncertainty on the branching ratio is lower than 1\% and is neglected. 

The total systematic uncertainty is calculated as the quadratic sum of the individual contributions.

\section{Results}
\begin{figure}[ht]
    \centering
    \includegraphics[width=0.98\textwidth]{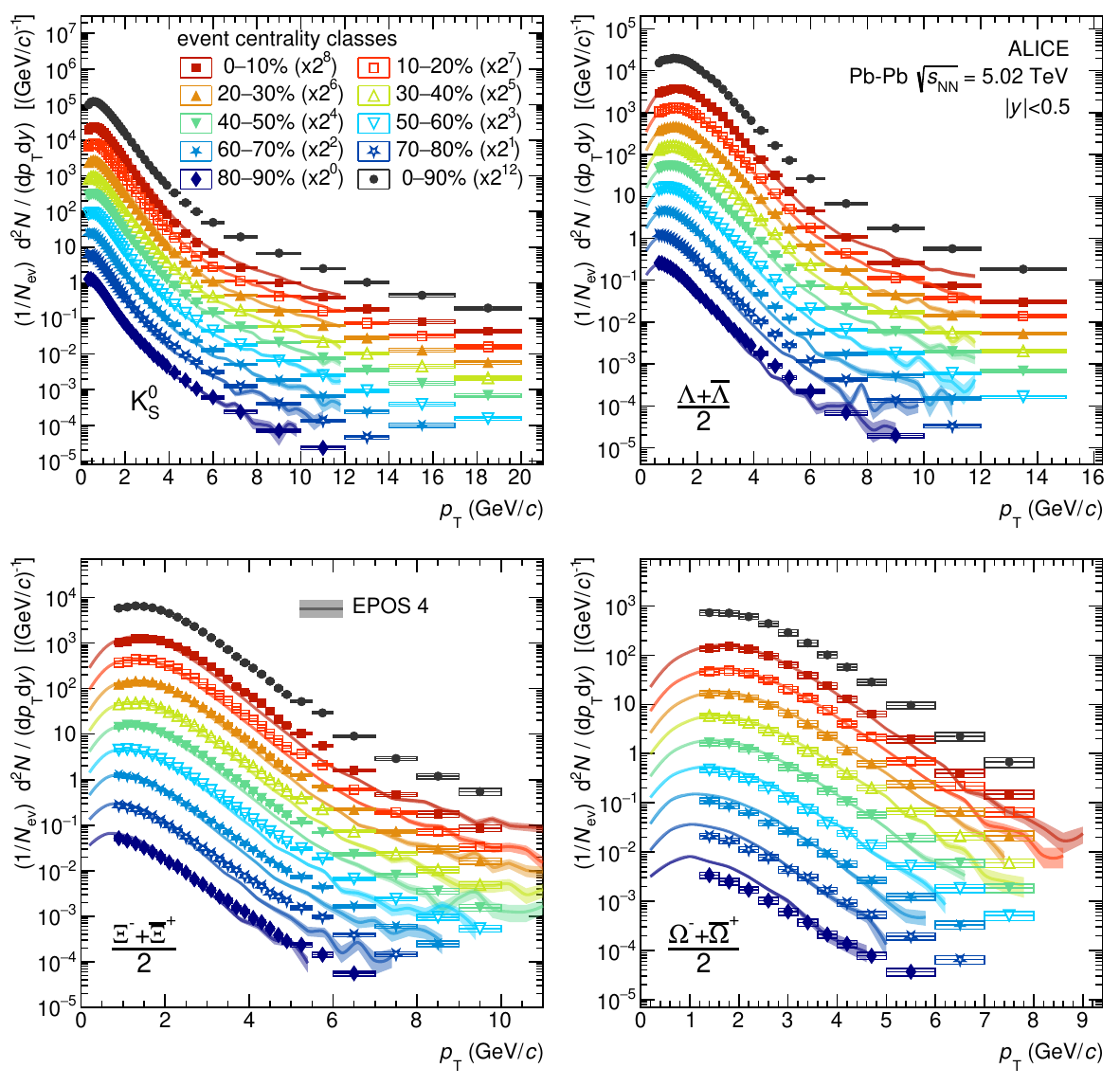}
     \caption{\kzero, \lmb, \Xit, and \Omt transverse momentum spectra in \PbPb collisions at \fivenn for different event centrality intervals. Error bars show the statistical uncertainty, while boxes represent the total systematic uncertainty. For each centrality class, predictions from the \epf~\cite{EPOS4:strangeness} model are reported as solid lines, with the corresponding transparent area accounting for the statistical uncertainty on the model. Scale factors are applied to both the results and the model to improve visibility.}
     \label{fig:spectra}
\end{figure}
\begin{figure}
    \centering
    \includegraphics[width=0.48 \textwidth]{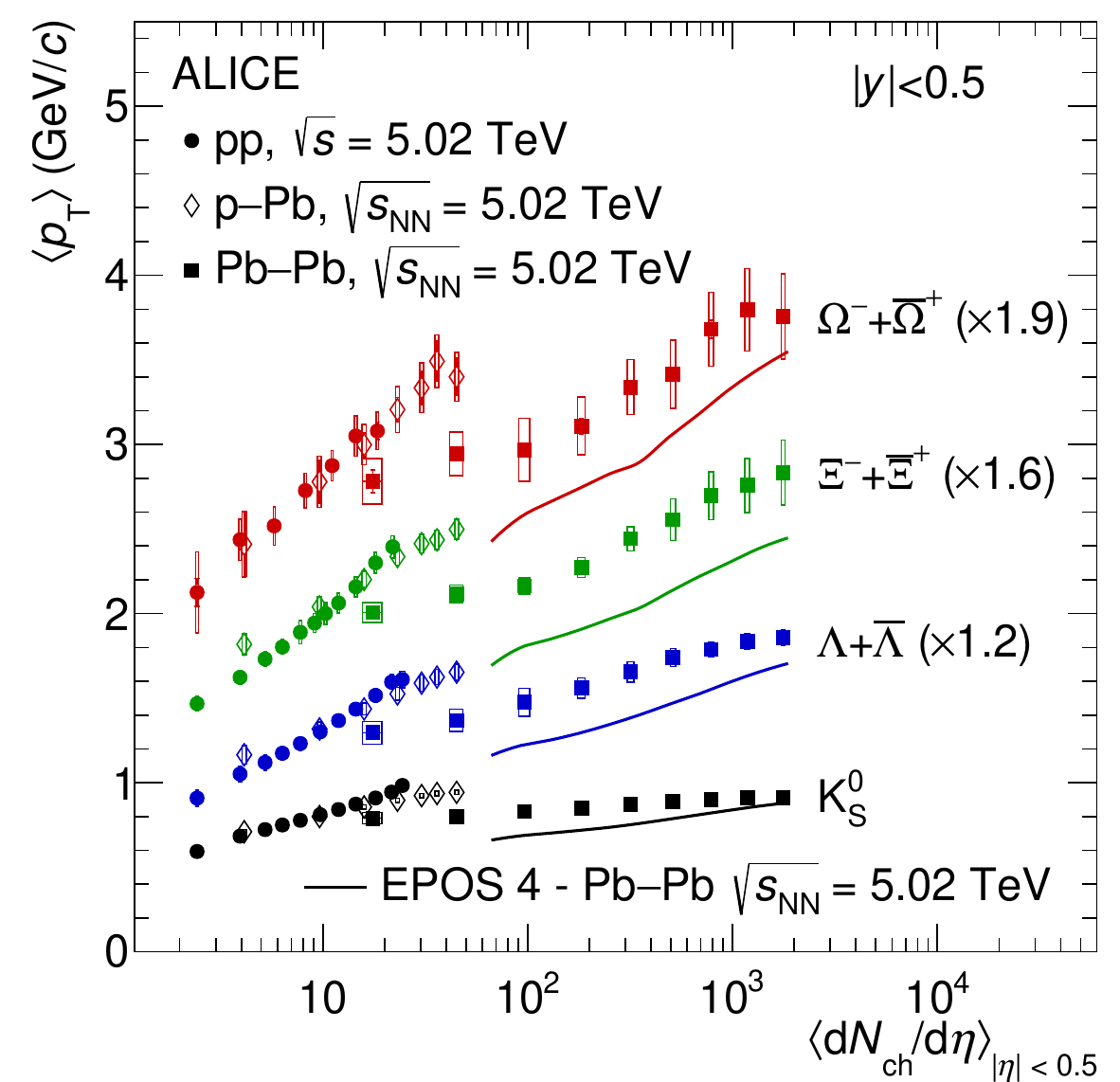} 
    \hspace*{0.4cm}
    \includegraphics[width=0.48 \textwidth]{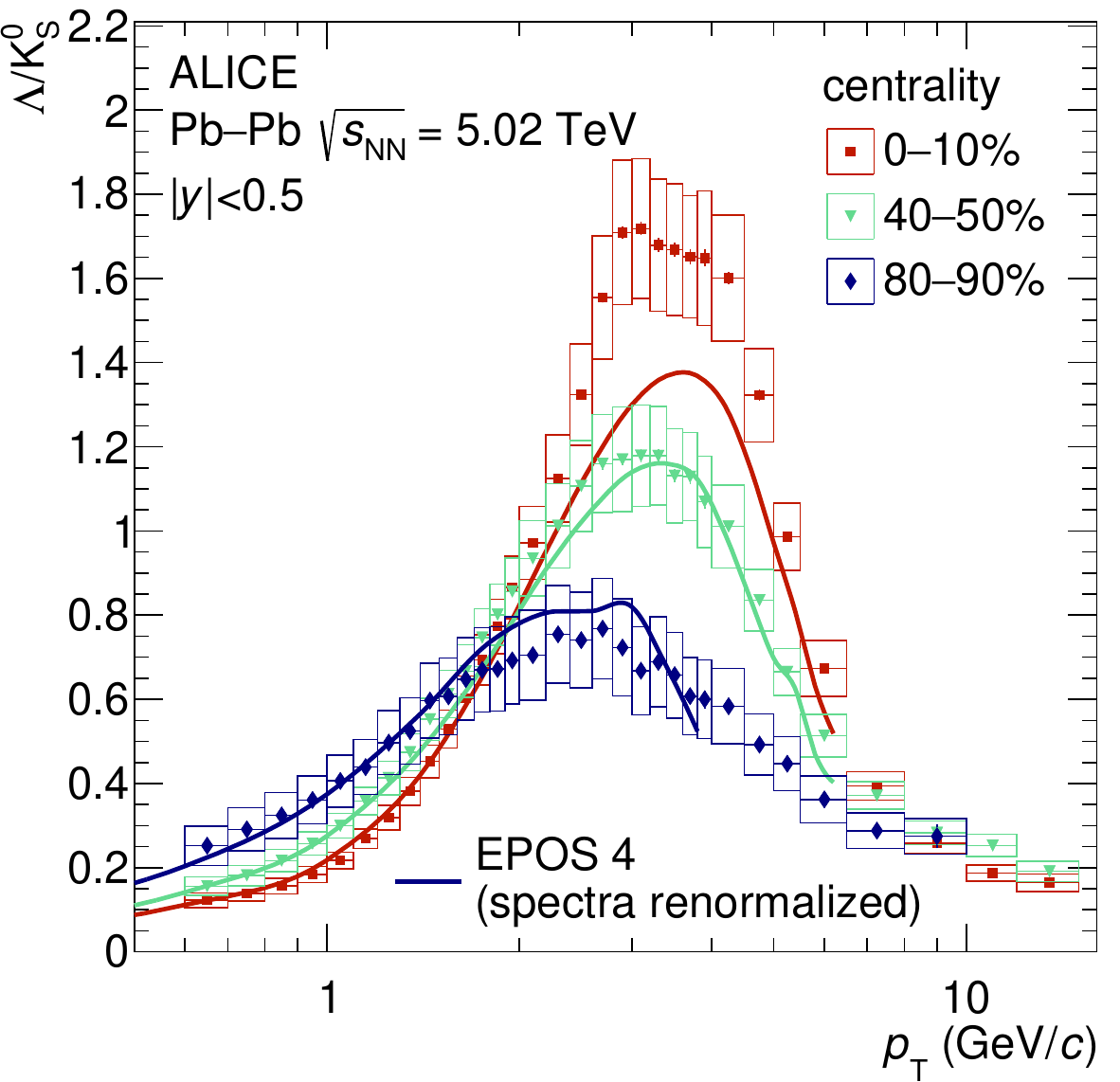}
    \caption{(left) \meanpt as a function of \avdndeta for \kzero, \lmb, \Xit, and \Omt particle production in \PbPb collisions at \fivenn at midrapidity (square markers). Results from \pp~\cite{silviapp5_TBU} and \pPb~\cite{Alice_strpPb5,ALICE_cascpPb5} collisions at the same center-of-mass energy are also reported as circles and open diamonds, respectively. Error bars correspond to the statistical uncertainty, while boxes correspond to the total systematic uncertainty. Solid lines are predictions from the \epf~\cite{EPOS4:strangeness} model. (right) \lmb/\kzero ratio as a function of \pt for the 0--10\% most central (red), 40--50\% intermediate (green) and 80--90\% most peripheral (blue) centrality classes. Error bars correspond to the statistical uncertainty, while boxes correspond to the total systematic uncertainty. \epf~\cite{EPOS4:strangeness} predictions, scaled to the total particle yields in the corresponding centrality class, are reported as solid lines.}
    \label{fig:SpectralShape}
\end{figure}

Transverse momentum spectra of particles and their charge conjugates are found to be compatible within uncertainties in all the \pt range, as expected at LHC energies and previously reported~\cite{Alice_pikppbpb276}. For this reason, all results presented here correspond to the average of particles and antiparticles. \\
The corrected \pt spectra of strange hadrons \kzero, \lmb, \Xit, and \Omt in \PbPb collisions at \fivenn are shown in Fig.~\ref{fig:spectra} for the different centrality classes considered in the analysis along with the centrality-integrated (0--90\%) results. Nine centrality classes are considered for all particles,  achieving high granularity in \pt and extending the momentum reach with respect to previous results, thanks to the large data samples acquired during the LHC Run 2 campaign. For all the measured particle species a hardening of the \pt spectrum moving from peripheral to central events is observed. The measured spectra are compared to predictions from the \epf event generator, which qualitatively reproduces the trend with centrality. \epf describes the \kzero and \lmb spectra within uncertainties for $\pt<4$~\GeVc in all centrality classes, while it underestimates the ones of \Xit at $\pt>2$~\GeVc and overestimates the \Omt spectrum in peripheral collisions. In this model the hardening of the spectra from peripheral to central collisions reflects the radial flow of partons in the hydrodynamically expanding colored medium. This effect can be studied more quantitatively by looking at the dependence of the \meanpt on the average charged particle multiplicity at midrapidity \multdensmid, calculated for each centrality class, as reported in Fig.~\ref{fig:SpectralShape}~(left) and in Table~\ref{tab:yieldmeanpt}. A clear multiplicity dependence is observed for all particles under study. This phenomenon has been previously reported in \pp~\cite{silviapp5_TBU} and \pPb~\cite{Alice_strpPb5,ALICE_cascpPb5} collisions at the same center-of-mass energy. The results from such systems are also reported in Fig.~\ref{fig:SpectralShape}~(left), where a clear discontinuity among the trends at high multiplicity in \pp, in \pPb and at low multiplicity \PbPb is observed, as previously reported for charged particles~\cite{Alice_avptppppbpbpb}. This suggests that similar multiplicities in different collision systems are achieved by means of softer interactions in large collision systems, or that re-scattering processes in \PbPb tend to decrease the \meanpt of the final state hadrons. The \meanpt in \epf is significantly lower than the one observed in the data, with a smaller deviation in more central collisions.

The \lmb/\kzero yield ratio as a function of the transverse momentum is an interesting observable, as it is sensitive to different competing physics processes in the strange baryon/meson formation: hydrodynamic expansion, quark recombination, and jet fragmentation at low, intermediate, and high \pt~\cite{ALICE_v0PbPb276}. Figure~\ref{fig:SpectralShape}~(right) reports this ratio in three selected centrality classes to improve visibility. The effect turns out to be progressive when considering all centrality classes (results corresponding to all centrality classes are available in the data compilation). In the high \pt region, the ratios from different centrality classes merge to a common value, a feature that is expected if particle production at high \pt is dominated by pQCD processes as jet fragmentation. At low \pt, a steep increase in the ratio is observed, more prominent in central than in peripheral collisions. In the 0--10\% centrality class the ratio reaches a peak of about 1.4 at \pt $\approx$ 3.5 \GeVc, while in the 80--90\% class it reaches a peak of around 0.6 at \pt $\approx$ 2.5 \GeVc. This observation is consistent with the hydrodynamic expansion picture, predicting that baryons should be pushed to higher \pt by radial flow more effectively than mesons. 
The phenomenological interpretation of this observable can be tested with the comparison to the \epf model, which includes all the physics mechanisms discussed and the cross-talk between hard and soft processes. As the objective here is to compare spectral-shape ratios, \epf spectra have been rescaled here to match the overall production yield for the considered particle in all centrality classes. The net-effect of the rescaling procedure is a $\approx17\%$ upwards shift of the \lmb/\kzero ratio for all centrality classes. The \pt reach of the model is restricted to 6 \GeVc due to limited statistics. The agreement with the model is very good in most central collisions, while the comparison worsens at intermediate and high \pt in semi-central and peripheral collisions, where EPOS predicts a larger radial boost than what is observed in the data.

\begin{table}[h]
\centering
\caption{\pt-integrated particle yields (\dndy) and \meanpt as a function of centrality for \kzero, \Lamboth, \Xiboth, and \Omboth hadron production in \PbPb collisions at \fivenn. The value $0.00$ indicates negligible statistical uncertainty.}
\label{tab:yieldmeanpt}
\renewcommand{\arraystretch}{1.2}
\begin{sideways}
\begin{tabular}{ccccc}
\hline
Centrality  & \kzero & \lmb & \Xit & \Omt \\
\hline
\multicolumn{5}{c}{\dndy$\pm$stat$\pm$syst} \\
\hline
0--10\%    & 115.69 $\pm$ 0.08 $\pm$ 7.96 & 55.62 $\pm$ 0.14 $\pm$ 4.26 & 10.14 $\pm$ 0.04 $\pm$ 0.99 & 1.36 $\pm$ 0.019 $\pm$ 0.234 \\
10--20\%   & 78.00 $\pm$ 0.05 $\pm$ 5.32 & 38.89 $\pm$ 0.10 $\pm$ 3.05 & 6.96 $\pm$ 0.02 $\pm$ 0.66 & 0.878 $\pm$ 0.016 $\pm$ 0.165 \\
20--30\%  & 52.91 $\pm$ 0.05 $\pm$ 3.97 & 27.17 $\pm$ 0.03 $\pm$ 2.13 & 4.66 $\pm$ 0.01 $\pm$ 0.41 & 0.601 $\pm$ 0.014 $\pm$ 0.109 \\
30--40\%   & 33.77  $\pm$ 0.03 $\pm$ 2.68 & 17.95 $\pm$ 0.03 $\pm$ 1.52 & 3.16 $\pm$ 0.01 $\pm$ 0.28 & 0.418 $\pm$ 0.009 $\pm$ 0.070 \\
40--50\% & 20.92  $\pm$ 0.08 $\pm$ 1.69 & 11.45 $\pm$ 0.02 $\pm$ 1.07 & 1.93 $\pm$ 0.01 $\pm$ 0.17 & 0.240 $\pm$ 0.011 $\pm$ 0.040 \\
50--60\%    & 11.93  $\pm$ 0.03 $\pm$ 1.09 & 6.72  $\pm$ 0.01 $\pm$ 0.69 & 1.13 $\pm$ 0.00 $\pm$ 0.11 & 0.133 $\pm$ 0.003 $\pm$ 0.022 \\
60--70\%   & 6.13  $\pm$ 0.01 $\pm$ 0.61 & 3.57  $\pm$ 0.01 $\pm$ 0.48 & 0.56 $\pm$ 0.00 $\pm$ 0.05 & 0.062 $\pm$ 0.002 $\pm$ 0.010 \\
70--80\%  & 2.75   $\pm$ 0.01 $\pm$ 0.32 & 1.70  $\pm$ 0.02 $\pm$ 0.24 & 0.23 $\pm$ 0.00 $\pm$ 0.03 & 0.024 $\pm$ 0.000 $\pm$ 0.004 \\
80--90\%    & 1.07   $\pm$ 0.00 $\pm$ 0.13 & 0.67  $\pm$ 0.00 $\pm$ 0.11 & 0.08 $\pm$ 0.00 $\pm$ 0.01 & 0.008 $\pm$ 0.000 $\pm$ 0.001 \\
0--90\% & 36.81 $\pm$ 0.05 $\pm$ 2.95 & 17.93 $\pm$ 0.03 $\pm$ 1.77 & 3.23 $\pm$ 0.00 $\pm$ 0.31 & 0.412 $\pm$ 0.004 $\pm$ 0.082 \\
\hline
\multicolumn{5}{c}{\meanpt$\pm$stat$\pm$syst (\GeVc)} \\
\hline
0--10\%    & 0.94 $\pm$ 0.00 $\pm$ 0.02  & 1.55 $\pm$ 0.00 $\pm$ 0.04 & 1.77 $\pm$ 0.00 $\pm$ 0.12 & 1.98 $\pm$ 0.02 $\pm$ 0.13 \\
10--20\%    & 0.94 $\pm$ 0.00 $\pm$ 0.02  & 1.53 $\pm$ 0.00 $\pm$ 0.04 & 1.72 $\pm$ 0.00 $\pm$ 0.10 & 2.00 $\pm$ 0.02 $\pm$ 0.13 \\
20--30\%   & 0.94 $\pm$ 0.00 $\pm$ 0.02  & 1.49 $\pm$ 0.00 $\pm$ 0.03 & 1.69 $\pm$ 0.00 $\pm$ 0.09 & 1.94 $\pm$ 0.03 $\pm$ 0.11 \\
30--40\%    & 0.94 $\pm$ 0.00 $\pm$ 0.02  & 1.45 $\pm$ 0.00 $\pm$ 0.04 & 1.60 $\pm$ 0.00 $\pm$ 0.08 & 1.80 $\pm$ 0.02 $\pm$ 0.11 \\
40--50\%     & 0.92 $\pm$ 0.00 $\pm$ 0.02  & 1.38 $\pm$ 0.00 $\pm$ 0.05 & 1.53 $\pm$ 0.00 $\pm$ 0.05 & 1.76 $\pm$ 0.02 $\pm$ 0.09 \\
50--60\%    & 0.90 $\pm$ 0.00 $\pm$ 0.03  & 1.30 $\pm$ 0.00 $\pm$ 0.05 & 1.42 $\pm$ 0.00 $\pm$ 0.04 & 1.64 $\pm$ 0.02 $\pm$ 0.09 \\
60--70\%   & 0.89 $\pm$ 0.00 $\pm$ 0.03  & 1.23 $\pm$ 0.00 $\pm$ 0.07 & 1.35 $\pm$ 0.00 $\pm$ 0.03 & 1.56 $\pm$ 0.01 $\pm$ 0.10 \\
70--80\%  & 0.88 $\pm$ 0.00 $\pm$ 0.04  & 1.14 $\pm$ 0.01 $\pm$ 0.06 & 1.32 $\pm$ 0.00 $\pm$ 0.03 & 1.55 $\pm$ 0.01 $\pm$ 0.07 \\
80--90\%    & 0.85 $\pm$ 0.00 $\pm$ 0.03  & 1.08 $\pm$ 0.00 $\pm$ 0.06 & 1.25 $\pm$ 0.01 $\pm$ 0.04 & 1.46 $\pm$ 0.04 $\pm$ 0.07 \\
0--90\%    & 0.92 $\pm$ 0.00 $\pm$ 0.03 & 1.50 $\pm$ 0.00 $\pm$ 0.05 & 1.67 $\pm$ 0.00 $\pm$ 0.12 & 1.92 $\pm$ 0.01 $\pm$ 0.12 \\
\hline
\end{tabular}
\end{sideways}
\end{table}

\begin{figure}[ht]
    \centering
    \includegraphics[width=0.8\textwidth]{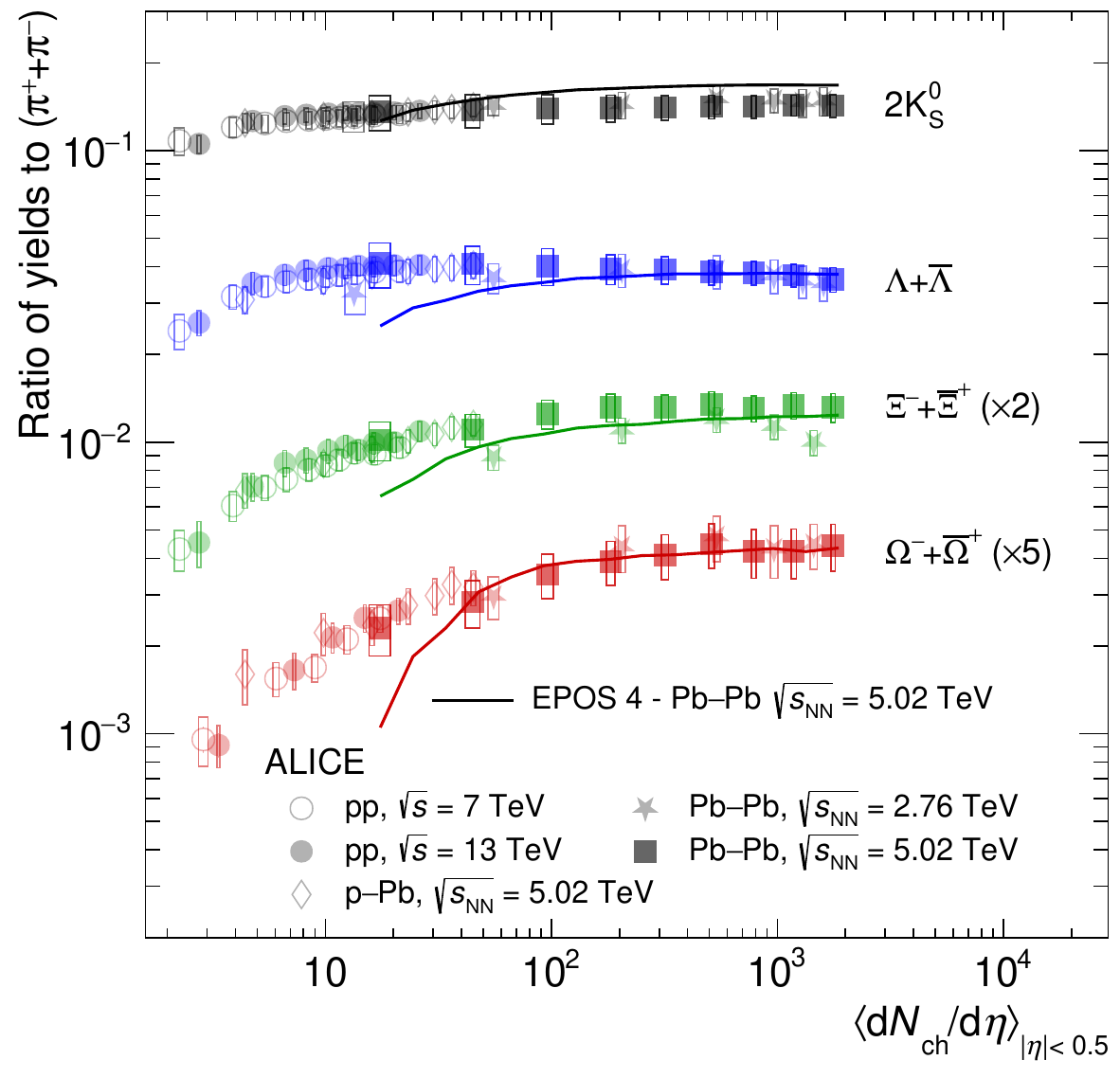}
     \caption{\kzero, \Lamboth, \Xiboth, and \Omboth particle yield ratios to the pion yield in \PbPb collisions at \fivenn (squares), compared to results in \pp collisions at \seven~\cite{ALICEstrPP7} and \thirteen~\cite{ALICEstrPP13} (hollow and full circles, respectively), in \pPb collisions at \five~\cite{ALICE_cascpPb5} (hollow diamonds), and in \PbPb collisions at \twosevensixnn~\cite{ALICE_v0PbPb276,ALICE_cascPbPb276} (stars). Error bars show the statistical uncertainty, while boxes represent the total systematic uncertainty. \epf~\cite{EPOS4:strangeness} predictions are reported for the top \PbPb energy as solid lines.}
     \label{fig:ToPionRatios}
\end{figure}

The \pt-integrated yields for all particles and centrality classes are reported in Table~\ref{tab:yieldmeanpt}. A distinct evolution of the yields with the collision centrality is observed, to some extent mapping the large variation in charged particle multiplicity across centrality. Thanks to the large data samples considered in the analysis, the statistical uncertainty is negligible as compared to the systematic one, not only for the abundant \kzero meson, but for cascades as well, notably for the \Omt.
To study how the probability of strange particle formation evolves relative to the activity in the analyzed events, the ratios of integrated yields between strange hadrons and pions~\cite{Alice_pikppbpb5} are evaluated and reported in Fig.~\ref{fig:ToPionRatios}. The plot also includes data measured by ALICE in \pp collisions at \seven~\cite{Alice_NPEnhProd,ALICEstrPP7} and \thirteen~\cite{ALICEstrPP13}, in \pPb collisions at \five~\cite{Alice_strpPb5,ALICE_cascpPb5}, and in \PbPb collisions at \twosevensixnn~\cite{ALICE_v0PbPb276,ALICE_cascPbPb276}.

The results show an increase of the strange-to-pion ratio in \PbPb compared to results from \pp and \pPb collisions. The enhancement is proportional to the strangeness content of the hadron, being the largest for cascades. The evolution of the ratio is progressive and smooth across multiplicity, perfectly superimposing for all collision systems in the common multiplicity region ($10\lesssim\avdndeta\lesssim100$). Focusing on the measurement in \PbPb collisions, there is no significant multiplicity dependence of the ratio for all particles under study, except for the \Omt/$\pi$ ratio. It has been proposed~\cite{SAVCHUK2022136983} that baryon annihilation in the hadronic phase after crossover could result in a depletion of the baryon/pion ratio in central collisions with respect to peripheral. This would be more visible for the \lmb baryon, as the $\pi-\lmb$ cross section is expected to be higher than the $\pi-\Xi$ and $\pi-\Omega$ ones~\cite{Bleicher:1999xi}. With the presented results, it is not possible to conclude on a significant evolution of this ratio, though a mild decreasing trend is observed in \lmb/$\pi$, well inside the systematic uncertainty.
Consistent with what has been previously observed in \pp, the collision center-of-mass energy has no impact on the strange hadron relative production rate once similar multiplicities at different energies are compared: results at \five and \twosevensixnn are in very good agreement, with one single data point at the lower energy lying outside the common trend by more than one standard deviation (central \Xit, $2.4\sigma$ from the results at higher collision energy).
All these observations suggest that, in the energy range probed at the LHC, charged-particle multiplicity density is a good scaling variable for strange hadron abundances. 

The evolution of the ratio is rather well reproduced by \epf, thus supporting the recent tuning of the core--corona mechanism which is responsible for the strangeness enhancement pattern in the model. \kzero are slightly overestimated, while the low-multiplicity tail of the trend for \lmb, \Xit, and \Omt tends to be underestimated by the model.

\section{Conclusions}
Results on the production of \kzero, \Lamboth, \Xiboth, and \Omboth, in Pb--Pb collisions at \fivenn are reported.
Thanks to the large data samples considered in the analysis, \pt spectra from all the above-mentioned particles have been extracted in nine centrality classes, with \multdensmid ranging over two orders of magnitude, from less than 20 to almost 2000. This enables probing (multi-)strange particle production in multiplicity regions that are typical for \pPb or high multiplicity \pp collisions, on one side, and yet unexplored experimentally on the higher end.

Transverse momentum spectra evolve with centrality, featuring a higher \meanpt in more central collisions. This is compatible with the expectations from a collective colored fluid that expands hydrodynamically, as shown in comparison to the \epf MC generator. The EPOS MC model nicely reproduces the spectra, though having a $\approx$15\% lower \meanpt across all centralities and for all particles. Spectra evolution has been studied also through the \lmb/\kzero ratio as a function of \pt. The ratio features a bump at intermediate \pt, with a maximum at \pt $\approx$ 2-5 GeV/c, whose magnitude is larger in central than in peripheral collisions. At the highest \pt the ratios are similar for all centralities, thus showing that spectra evolution is limited to the low- and intermediate-\pt (non-perturbative) region. The comparison to the \epf model shows a good agreement in central collisions, worsening in semi-central and peripheral ones, where the MC prediction overestimates the effect at the peak. Nonetheless, the low-\pt hydrodynamic-like increasing trend is well reproduced by the model.

Production yields have been extracted for all particles in every centrality class by integrating the \pt spectra and extrapolating in the unmeasured region. The ratio between the yields of strange particles and those of pions shows a very limited evolution with centrality, with an hint of an increasing trend, not fully significant within the current systematic uncertainties, for the \Omboth baryon. For the \lmb baryon, a hint of decrease with multiplicity is observed, though not significant given the present systematic uncertainties. Yield ratios have been compared to the same quantity extracted at lower energy and for smaller collision systems at different energies. The evolution of particle yield ratios with multiplicity, reported from $\approx$2 to $\sim$2000 average charged-particle density at midrapidity, follows a common pattern, which does not depend on the center-of-mass energy of the collision nor on the collision system type. 
The comparison of the yield ratio to pions with the \epf model shows a good agreement, with a depletion towards low multiplicity that is more pronounced in the model than in the data.

\section*{Data availability statement}
This manuscript has associated data in a HEPData repository at Ref.~\cite{hepdata.180087}.

\newenvironment{acknowledgement}{\relax}{\relax}
\begin{acknowledgement}
\section*{Acknowledgements}

The ALICE Collaboration would like to thank all its engineers and technicians for their invaluable contributions to the construction of the experiment and the CERN accelerator teams for the outstanding performance of the LHC complex.
The ALICE Collaboration gratefully acknowledges the resources and support provided by all Grid centres and the Worldwide LHC Computing Grid (WLCG) collaboration.
The ALICE Collaboration acknowledges the following funding agencies for their support in building and running the ALICE detector:
A. I. Alikhanyan National Science Laboratory (Yerevan Physics Institute) Foundation (ANSL), State Committee of Science and World Federation of Scientists (WFS), Armenia;
Austrian Academy of Sciences, Austrian Science Fund (FWF): [M 2467-N36] and Nationalstiftung f\"{u}r Forschung, Technologie und Entwicklung, Austria;
Ministry of Communications and High Technologies, National Nuclear Research Center, Azerbaijan;
Rede Nacional de Física de Altas Energias (Renafae), Financiadora de Estudos e Projetos (Finep), Funda\c{c}\~{a}o de Amparo \`{a} Pesquisa do Estado de S\~{a}o Paulo (FAPESP) and The Sao Paulo Research Foundation  (FAPESP), Brazil;
Bulgarian Ministry of Education and Science, within the National Roadmap for Research Infrastructures 2020-2027 (object CERN), Bulgaria;
Ministry of Education of China (MOEC) , Ministry of Science \& Technology of China (MSTC) and National Natural Science Foundation of China (NSFC), China;
Ministry of Science and Education and Croatian Science Foundation, Croatia;
Centro de Aplicaciones Tecnol\'{o}gicas y Desarrollo Nuclear (CEADEN), Cubaenerg\'{\i}a, Cuba;
Ministry of Education, Youth and Sports of the Czech Republic, Czech Republic;
The Danish Council for Independent Research | Natural Sciences, the VILLUM FONDEN and Danish National Research Foundation (DNRF), Denmark;
Helsinki Institute of Physics (HIP), Finland;
Commissariat \`{a} l'Energie Atomique (CEA) and Institut National de Physique Nucl\'{e}aire et de Physique des Particules (IN2P3) and Centre National de la Recherche Scientifique (CNRS), France;
Bundesministerium f\"{u}r Forschung, Technologie und Raumfahrt (BMFTR) and GSI Helmholtzzentrum f\"{u}r Schwerionenforschung GmbH, Germany;
General Secretariat for Research and Technology, Ministry of Education, Research and Religions, Greece;
National Research, Development and Innovation Office, Hungary;
Department of Atomic Energy Government of India (DAE), Department of Science and Technology, Government of India (DST), University Grants Commission, Government of India (UGC) and Council of Scientific and Industrial Research (CSIR), India;
National Research and Innovation Agency - BRIN, Indonesia;
Istituto Nazionale di Fisica Nucleare (INFN), Italy;
Japanese Ministry of Education, Culture, Sports, Science and Technology (MEXT) and Japan Society for the Promotion of Science (JSPS) KAKENHI, Japan;
Consejo Nacional de Ciencia (CONACYT) y Tecnolog\'{i}a, through Fondo de Cooperaci\'{o}n Internacional en Ciencia y Tecnolog\'{i}a (FONCICYT) and Direcci\'{o}n General de Asuntos del Personal Academico (DGAPA), Mexico;
Nederlandse Organisatie voor Wetenschappelijk Onderzoek (NWO), Netherlands;
The Research Council of Norway, Norway;
Pontificia Universidad Cat\'{o}lica del Per\'{u}, Peru;
Ministry of Science and Higher Education, National Science Centre and WUT ID-UB, Poland;
Korea Institute of Science and Technology Information and National Research Foundation of Korea (NRF), Republic of Korea;
Ministry of Education and Scientific Research, Institute of Atomic Physics, Ministry of Research and Innovation and Institute of Atomic Physics and Universitatea Nationala de Stiinta si Tehnologie Politehnica Bucuresti, Romania;
Ministerstvo skolstva, vyskumu, vyvoja a mladeze SR, Slovakia;
National Research Foundation of South Africa, South Africa;
Swedish Research Council (VR) and Knut \& Alice Wallenberg Foundation (KAW), Sweden;
European Organization for Nuclear Research, Switzerland;
Suranaree University of Technology (SUT), National Science and Technology Development Agency (NSTDA) and National Science, Research and Innovation Fund (NSRF via PMU-B B05F650021), Thailand;
Turkish Energy, Nuclear and Mineral Research Agency (TENMAK), Turkey;
National Academy of  Sciences of Ukraine, Ukraine;
Science and Technology Facilities Council (STFC), United Kingdom;
National Science Foundation of the United States of America (NSF) and United States Department of Energy, Office of Nuclear Physics (DOE NP), United States of America.
In addition, individual groups or members have received support from:
Czech Science Foundation (grant no. 23-07499S), Czech Republic;
FORTE project, reg.\ no.\ CZ.02.01.01/00/22\_008/0004632, Czech Republic, co-funded by the European Union, Czech Republic;
European Research Council (grant no. 950692), European Union;
Deutsche Forschungs Gemeinschaft (DFG, German Research Foundation) ``Neutrinos and Dark Matter in Astro- and Particle Physics'' (grant no. SFB 1258), Germany;
CONVECS project, CUP C97H23001700002 FESR 2021-2027 program, Italy;
FAIR - Future Artificial Intelligence Research, funded by the NextGenerationEU program (Italy).

\end{acknowledgement}

\bibliographystyle{utphys}
\bibliography{bibliography}

\newpage
\appendix

\section{The ALICE Collaboration}
\label{app:collab}
\begin{flushleft} 
\small

I.J.~Abualrob\,\orcidlink{0009-0005-3519-5631}\,$^{\rm 114}$, 
S.~Acharya\,\orcidlink{0000-0002-9213-5329}\,$^{\rm 50}$, 
G.~Aglieri Rinella\,\orcidlink{0000-0002-9611-3696}\,$^{\rm 32}$, 
L.~Aglietta\,\orcidlink{0009-0003-0763-6802}\,$^{\rm 24}$, 
N.~Agrawal\,\orcidlink{0000-0003-0348-9836}\,$^{\rm 25}$, 
Z.~Ahammed\,\orcidlink{0000-0001-5241-7412}\,$^{\rm 134}$, 
S.~Ahmad\,\orcidlink{0000-0003-0497-5705}\,$^{\rm 15}$, 
I.~Ahuja\,\orcidlink{0000-0002-4417-1392}\,$^{\rm 36}$, 
Z.~Akbar$^{\rm 81}$, 
A.~Akindinov\,\orcidlink{0000-0002-7388-3022}\,$^{\rm 140}$
% ...

\section*{Affiliation Notes}

$^{\rm I}$ Deceased\\
$^{\rm II}$ Also at: Max-Planck-Institut fur Physik, Munich, Germany\\
$^{\rm III}$ Also at: Czech Technical University in Prague (CZ)\\
$^{\rm IV}$ Also at: Instituto de Fisica da Universidade de Sao Paulo\\
$^{\rm V}$ Also at: Dipartimento DET del Politecnico di Torino, Turin, Italy\\
$^{\rm VI}$ Also at: Department of Applied Physics, Aligarh Muslim University, Aligarh, India\\
$^{\rm VII}$ Also at: Institute of Theoretical Physics, University of Wroclaw, Poland\\
$^{\rm VIII}$ Also at: Facultad de Ciencias, Universidad Nacional Aut\'{o}noma de M\'{e}xico, Mexico City, Mexico\\
$^{\rm IX}$ Also at: University College of Dublin, Dublin, Ireland\\

\section*{Collaboration Institutes}

$^{1}$ A.I. Alikhanyan National Science Laboratory (Yerevan Physics Institute) Foundation, Yerevan, Armenia\\
$^{2}$ AGH University of Krakow, Cracow, Poland\\
$^{3}$ Bogolyubov Institute for Theoretical Physics, National Academy of Sciences of Ukraine, Kyiv, Ukraine\\
$^{4}$ Bose Institute, Department of Physics  and Centre for Astroparticle Physics and Space Science (CAPSS), Kolkata, India\\
$^{5}$ California Polytechnic State University, San Luis Obispo, California, United States\\
$^{6}$ Central China Normal University, Wuhan, China\\
$^{7}$ Centro de Aplicaciones Tecnol\'{o}gicas y Desarrollo Nuclear (CEADEN), Havana, Cuba\\
$^{8}$ Centro de Investigaci\'{o}n y de Estudios Avanzados (CINVESTAV), Mexico City and M\'{e}rida, Mexico\\
$^{9}$ Chicago State University, Chicago, Illinois, United States\\
$^{10}$ China Nuclear Data Center, China Institute of Atomic Energy, Beijing, China\\
$^{11}$ China University of Geosciences, Wuhan, China\\
$^{12}$ Chungbuk National University, Cheongju, Republic of Korea\\
$^{13}$ Comenius University Bratislava, Faculty of Mathematics, Physics and Informatics, Bratislava, Slovak Republic\\
$^{14}$ Creighton University, Omaha, Nebraska, United States\\
$^{15}$ Department of Physics, Aligarh Muslim University, Aligarh, India\\
$^{16}$ Department of Physics, Pusan National University, Pusan, Republic of Korea\\
$^{17}$ Department of Physics, Sejong University, Seoul, Republic of Korea\\
$^{18}$ Department of Physics, University of California, Berkeley, California, United States\\
$^{19}$ Department of Physics, University of Oslo, Oslo, Norway\\
$^{20}$ Department of Physics and Technology, University of Bergen, Bergen, Norway\\
$^{21}$ Dipartimento di Fisica, Universit\`{a} di Pavia, Pavia, Italy\\
$^{22}$ Dipartimento di Fisica dell'Universit\`{a} and Sezione INFN, Cagliari, Italy\\
$^{23}$ Dipartimento di Fisica dell'Universit\`{a} and Sezione INFN, Trieste, Italy\\
$^{24}$ Dipartimento di Fisica dell'Universit\`{a} and Sezione INFN, Turin, Italy\\
$^{25}$ Dipartimento di Fisica e Astronomia dell'Universit\`{a} and Sezione INFN, Bologna, Italy\\
$^{26}$ Dipartimento di Fisica e Astronomia dell'Universit\`{a} and Sezione INFN, Catania, Italy\\
$^{27}$ Dipartimento di Fisica e Astronomia dell'Universit\`{a} and Sezione INFN, Padova, Italy\\
$^{28}$ Dipartimento di Fisica `E.R.~Caianiello' dell'Universit\`{a} and Gruppo Collegato INFN, Salerno, Italy\\
$^{29}$ Dipartimento DISAT del Politecnico and Sezione INFN, Turin, Italy\\
$^{30}$ Dipartimento di Scienze MIFT, Universit\`{a} di Messina, Messina, Italy\\
$^{31}$ Dipartimento Interateneo di Fisica `M.~Merlin' and Sezione INFN, Bari, Italy\\
$^{32}$ European Organization for Nuclear Research (CERN), Geneva, Switzerland\\
$^{33}$ Faculty of Electrical Engineering, Mechanical Engineering and Naval Architecture, University of Split, Split, Croatia\\
$^{34}$ Faculty of Nuclear Sciences and Physical Engineering, Czech Technical University in Prague, Prague, Czech Republic\\
$^{35}$ Faculty of Physics, Sofia University, Sofia, Bulgaria\\
$^{36}$ Faculty of Science, P.J.~\v{S}af\'{a}rik University, Ko\v{s}ice, Slovak Republic\\
$^{37}$ Faculty of Technology, Environmental and Social Sciences, Bergen, Norway\\
$^{38}$ Frankfurt Institute for Advanced Studies, Johann Wolfgang Goethe-Universit\"{a}t Frankfurt, Frankfurt, Germany\\
$^{39}$ Fudan University, Shanghai, China\\
$^{40}$ Gangneung-Wonju National University, Gangneung, Republic of Korea\\
$^{41}$ Gauhati University, Department of Physics, Guwahati, India\\
$^{42}$ Helmholtz-Institut f\"{u}r Strahlen- und Kernphysik, Rheinische Friedrich-Wilhelms-Universit\"{a}t Bonn, Bonn, Germany\\
$^{43}$ Helsinki Institute of Physics (HIP), Helsinki, Finland\\
$^{44}$ High Energy Physics Group,  Universidad Aut\'{o}noma de Puebla, Puebla, Mexico\\
$^{45}$ Horia Hulubei National Institute of Physics and Nuclear Engineering, Bucharest, Romania\\
$^{46}$ HUN-REN Wigner Research Centre for Physics, Budapest, Hungary\\
$^{47}$ Indian Institute of Technology Bombay (IIT), Mumbai, India\\
$^{48}$ Indian Institute of Technology Indore, Indore, India\\
$^{49}$ INFN, Laboratori Nazionali di Frascati, Frascati, Italy\\
$^{50}$ INFN, Sezione di Bari, Bari, Italy\\
$^{51}$ INFN, Sezione di Bologna, Bologna, Italy\\
$^{52}$ INFN, Sezione di Cagliari, Cagliari, Italy\\
$^{53}$ INFN, Sezione di Catania, Catania, Italy\\
$^{54}$ INFN, Sezione di Padova, Padova, Italy\\
$^{55}$ INFN, Sezione di Pavia, Pavia, Italy\\
$^{56}$ INFN, Sezione di Torino, Turin, Italy\\
$^{57}$ INFN, Sezione di Trieste, Trieste, Italy\\
$^{58}$ Inha University, Incheon, Republic of Korea\\
$^{59}$ Institute for Gravitational and Subatomic Physics (GRASP), Utrecht University/Nikhef, Utrecht, Netherlands\\
$^{60}$ Institute of Experimental Physics, Slovak Academy of Sciences, Ko\v{s}ice, Slovak Republic\\
$^{61}$ Institute of Physics, Homi Bhabha National Institute, Bhubaneswar, India\\
$^{62}$ Institute of Physics of the Czech Academy of Sciences, Prague, Czech Republic\\
$^{63}$ Institute of Space Science (ISS), Bucharest, Romania\\
$^{64}$ Institut f\"{u}r Kernphysik, Johann Wolfgang Goethe-Universit\"{a}t Frankfurt, Frankfurt, Germany\\
$^{65}$ Instituto de Ciencias Nucleares, Universidad Nacional Aut\'{o}noma de M\'{e}xico, Mexico City, Mexico\\
$^{66}$ Instituto de F\'{i}sica, Universidade Federal do Rio Grande do Sul (UFRGS), Porto Alegre, Brazil\\
$^{67}$ Instituto de F\'{\i}sica, Universidad Nacional Aut\'{o}noma de M\'{e}xico, Mexico City, Mexico\\
$^{68}$ iThemba LABS, National Research Foundation, Somerset West, South Africa\\
$^{69}$ Jeonbuk National University, Jeonju, Republic of Korea\\
$^{70}$ Korea Institute of Science and Technology Information, Daejeon, Republic of Korea\\
$^{71}$ Laboratoire de Physique Subatomique et de Cosmologie, Universit\'{e} Grenoble-Alpes, CNRS-IN2P3, Grenoble, France\\
$^{72}$ Lawrence Berkeley National Laboratory, Berkeley, California, United States\\
$^{73}$ Lund University Department of Physics, Division of Particle Physics, Lund, Sweden\\
$^{74}$ Marietta Blau Institute, Vienna, Austria\\
$^{75}$ Nagasaki Institute of Applied Science, Nagasaki, Japan\\
$^{76}$ Nara Women{'}s University (NWU), Nara, Japan\\
$^{77}$ National and Kapodistrian University of Athens, School of Science, Department of Physics , Athens, Greece\\
$^{78}$ National Centre for Nuclear Research, Warsaw, Poland\\
$^{79}$ National Institute of Science Education and Research, Homi Bhabha National Institute, Jatni, India\\
$^{80}$ National Nuclear Research Center, Baku, Azerbaijan\\
$^{81}$ National Research and Innovation Agency - BRIN, Jakarta, Indonesia\\
$^{82}$ Niels Bohr Institute, University of Copenhagen, Copenhagen, Denmark\\
$^{83}$ Nikhef, National institute for subatomic physics, Amsterdam, Netherlands\\
$^{84}$ Nuclear Physics Group, STFC Daresbury Laboratory, Daresbury, United Kingdom\\
$^{85}$ Nuclear Physics Institute of the Czech Academy of Sciences, Husinec-\v{R}e\v{z}, Czech Republic\\
$^{86}$ Oak Ridge National Laboratory, Oak Ridge, Tennessee, United States\\
$^{87}$ Ohio State University, Columbus, Ohio, United States\\
$^{88}$ Physics department, Faculty of science, University of Zagreb, Zagreb, Croatia\\
$^{89}$ Physics Department, Panjab University, Chandigarh, India\\
$^{90}$ Physics Department, University of Jammu, Jammu, India\\
$^{91}$ Physics Program and International Institute for Sustainability with Knotted Chiral Meta Matter (WPI-SKCM$^{2}$), Hiroshima University, Hiroshima, Japan\\
$^{92}$ Physikalisches Institut, Eberhard-Karls-Universit\"{a}t T\"{u}bingen, T\"{u}bingen, Germany\\
$^{93}$ Physikalisches Institut, Ruprecht-Karls-Universit\"{a}t Heidelberg, Heidelberg, Germany\\
$^{94}$ Physik Department, Technische Universit\"{a}t M\"{u}nchen, Munich, Germany\\
$^{95}$ Politecnico di Bari and Sezione INFN, Bari, Italy\\
$^{96}$ Research Division and ExtreMe Matter Institute EMMI, GSI Helmholtzzentrum f\"ur Schwerionenforschung GmbH, Darmstadt, Germany\\
$^{97}$ Saga University, Saga, Japan\\
$^{98}$ Saha Institute of Nuclear Physics, Homi Bhabha National Institute, Kolkata, India\\
$^{99}$ School of Physics and Astronomy, University of Birmingham, Birmingham, United Kingdom\\
$^{100}$ Secci\'{o}n F\'{\i}sica, Departamento de Ciencias, Pontificia Universidad Cat\'{o}lica del Per\'{u}, Lima, Peru\\
$^{101}$ SUBATECH, IMT Atlantique, Nantes Universit\'{e}, CNRS-IN2P3, Nantes, France\\
$^{102}$ Sungkyunkwan University, Suwon City, Republic of Korea\\
$^{103}$ Suranaree University of Technology, Nakhon Ratchasima, Thailand\\
$^{104}$ Technical University of Ko\v{s}ice, Ko\v{s}ice, Slovak Republic\\
$^{105}$ The Henryk Niewodniczanski Institute of Nuclear Physics, Polish Academy of Sciences, Cracow, Poland\\
$^{106}$ The University of Texas at Austin, Austin, Texas, United States\\
$^{107}$ Universidad Aut\'{o}noma de Sinaloa, Culiac\'{a}n, Mexico\\
$^{108}$ Universidade de S\~{a}o Paulo (USP), S\~{a}o Paulo, Brazil\\
$^{109}$ Universidade Estadual de Campinas (UNICAMP), Campinas, Brazil\\
$^{110}$ Universidade Federal do ABC, Santo Andre, Brazil\\
$^{111}$ Universitatea Nationala de Stiinta si Tehnologie Politehnica Bucuresti, Bucharest, Romania\\
$^{112}$ University of Cape Town, Cape Town, South Africa\\
$^{113}$ University of Derby, Derby, United Kingdom\\
$^{114}$ University of Houston, Houston, Texas, United States\\
$^{115}$ University of Jyv\"{a}skyl\"{a}, Jyv\"{a}skyl\"{a}, Finland\\
$^{116}$ University of Kansas, Lawrence, Kansas, United States\\
$^{117}$ University of Liverpool, Liverpool, United Kingdom\\
$^{118}$ University of Science and Technology of China, Hefei, China\\
$^{119}$ University of Silesia in Katowice, Katowice, Poland\\
$^{120}$ University of South-Eastern Norway, Kongsberg, Norway\\
$^{121}$ University of Tennessee, Knoxville, Tennessee, United States\\
$^{122}$ University of the Witwatersrand, Johannesburg, South Africa\\
$^{123}$ University of Tokyo, Tokyo, Japan\\
$^{124}$ University of Tsukuba, Tsukuba, Japan\\
$^{125}$ Universit\"{a}t M\"{u}nster, Institut f\"{u}r Kernphysik, M\"{u}nster, Germany\\
$^{126}$ Universit\'{e} Clermont Auvergne, CNRS/IN2P3, LPC, Clermont-Ferrand, France\\
$^{127}$ Universit\'{e} de Lyon, CNRS/IN2P3, Institut de Physique des 2 Infinis de Lyon, Lyon, France\\
$^{128}$ Universit\'{e} de Strasbourg, CNRS, IPHC UMR 7178, F-67000 Strasbourg, France, Strasbourg, France\\
$^{129}$ Universit\'{e} Paris-Saclay, Centre d'Etudes de Saclay (CEA), IRFU, D\'{e}partment de Physique Nucl\'{e}aire (DPhN), Saclay, France\\
$^{130}$ Universit\'{e}  Paris-Saclay, CNRS/IN2P3, IJCLab, Orsay, France\\
$^{131}$ Universit\`{a} degli Studi di Foggia, Foggia, Italy\\
$^{132}$ Universit\`{a} del Piemonte Orientale, Vercelli, Italy\\
$^{133}$ Universit\`{a} di Brescia, Brescia, Italy\\
$^{134}$ Variable Energy Cyclotron Centre, Homi Bhabha National Institute, Kolkata, India\\
$^{135}$ Warsaw University of Technology, Warsaw, Poland\\
$^{136}$ Wayne State University, Detroit, Michigan, United States\\
$^{137}$ Yale University, New Haven, Connecticut, United States\\
$^{138}$ Yildiz Technical University, Istanbul, Turkey\\
$^{139}$ Yonsei University, Seoul, Republic of Korea\\
$^{140}$ Affiliated with an institute formerly covered by a cooperation agreement with CERN\\
$^{141}$ Affiliated with an international laboratory covered by a cooperation agreement with CERN.\\

\end{flushleft}

\end{document}